\RequirePackage{amsmath}
\documentclass[runningheads]{llncs}
\usepackage[T1]{fontenc}
\usepackage{amssymb}
\usepackage{listings}
\newcommand{\code}[1]{{\footnotesize\texttt{#1}}}

\newcommand{\bi}{\begin{itemize}}
\newcommand{\ei}{\end{itemize}}
\begin{document}

\title{OpenBSD formal driver verification with SeL4}

% \author{}   % double-blind review
% \institute{}
% \author*[1]{\fnm{Adriana} \sur{Stancu}}\email{adriana.stancu@unibuc.ro}

% \author[1]{\fnm{Paul} \sur{Irofti}}\email{paul@irofti.net}

% \author[1]{\fnm{Ioana} \sur{Leuștean}}\email{ioana@fmi.unibuc.ro}
% %\equalcont{These authors contributed equally to this work.}

% \affil*[1]{\orgdiv{Computer Science}, \orgname{University of Bucharest}, \orgaddress{\street{Academiei 14}, \city{Bucharest}, \postcode{010014}, \country{Romania}}}

\author{
Adriana Nicolae\inst{1}
\and
Paul Irofti\inst{1,2}
\and
Ioana Leuștean\inst{1}
}
\institute{LOS-CS-FMI, University of Bucharest, Romania \and
Institute for Logic and Data Science, Romania\\
\email{adriana.stancu@unibuc.ro}, \email{paul@irofti.net}, \email{ioana@fmi.unibuc.ro}}

\maketitle
%--------------------------------------------------------------------------------------
\begin{abstract}
The seL4 microkernel is currently the only kernel that has been fully formally verified. In general, the increased interest in ensuring the security of a kernel's code results from its important role in the entire operating system.
One of the basic features of an operating system is that it abstracts the handling of devices. This abstraction is represented by device drivers - the software that manages the hardware. A proper verification of the software component could ensure that the device would work properly unless there is a hardware failure.
In this paper,
we choose to model the behavior of a device driver and build the proof that the code implementation matches the expected behavior.
The proof was written in Isabelle/HOL, the code translation from C to Isabelle was done automatically by the use of the C-to-Isabelle Parser and AutoCorres tools.
We choose Isabelle theorem prover because its efficiency was already shown through the verification of seL4 microkernel.

\end{abstract}
%--------------------------------------------------------------------------------------
\section{Introduction}

The kernel is a crucial component of the system, and direct access to hardware resources leads to an increased risk if a malfunction occurs. In our case, seL4 was designed as a microkernel in order to reduce the impact of software problems to the system's functionalities.

The main topic of interest in the analysis of the seL4 microkernel is the way to prove the functional correctness through the Isabelle/HOL theorem prover. The methods applied in system verification are more powerful and accurate than automated verification techniques such as model checking, static analysis, or deploying the entire kernel in a type-safe language.
This method of proving in Isabelle all the critical properties of the systems allows the analysis of specific aspects such as exploring the branches of execution of safe scenarios (safe execution), but also a set of specifications and proofs of kernel behavior reaching the analysis of implementation in C of the kernel for the ARM platform.

In this paper we investigate the process of adapting and applying the seL4 verification process for verify parts of another operating system and present a concrete case for the \code{octrng(4)} driver for the Octeon/MIPS64 platform provided by the OpenBSD operating system. 

\textbf{Outline.} In Section 2 we introduce the necessary seL4 concepts
which, together with the methodology from Section 3 regarding the translation of C code to Isabelle theorem prover,
allow us to present the verification of the OpenBSD driver in Section 4.
In Section 5 we conclude with limitations and future research.

%--------------------------------------------------------------------------------------
\section{SeL4 verification structure}
SeL4~\cite{sel4} is part of the L4 family - along with other implementations that share the same L4 interface: Pistachio \cite{pistachio}, Fiasco \cite{fiasco} or Hazelnut \cite{hazelnut}.
The proofs that underlie the verification of seL4 system are in the form of Hoare structures that have in their center a code component or whole functions. The difficulty of verifying the seL4 micro-kernel lies in formulating pre-conditions and post-conditions that accurately represent security properties that it must meet. At the same time, the formal representation must be as close as possible to the structure and functionalities implemented in the source code. Although work to implement its proofs was started in 2009 
%by the National ICT Australia - NICTA (Information and Communication Technology Research Center) research team,
formal proofs of the kernel are still maintained up to date with publicly available source code.

A key aspect of the design of a microkernel and the properties of the C code in relation to the form of their verification is the separation of kernel functions calls in two phases~\cite{verifos}: verification and execution. The verification phase can be understood as a stage of validation of the preconditions: the input data and the permissions on the actions to be performed are verified. The execution consists in the actual running of the system function, benefiting from its verification because the preconditions have already been verified in the previous phase.

Note that in the verification phase the system status is not changed otherwise this separation would no longer be relevant. This brings a valuable advantage in the verification process because it simplifies the system call proof: execution will not return an error if the verification phase has been completed successfully.

%--------------------------------------------------------------------------------------
\subsection{SeL4 memory management}
SeL4 kernel memory allocation model transfers allocation control from kernel space to applications that have this permission. Memory management permission is represented by having a structure called capability~\cite{sel4-wp}.
As a consequence the kernel heap memory can be precisely partitioned between applications: each application has that part of the heap for which it has a capability that gives it that authority. Separating heap memory is especially important for expressing and demonstrating security properties (integrity and confidentiality).

The basic features of the kernel memory allocation model are as follows~\cite{verifos}:
allocation is explicit and is performed only when assigning a type (retype) to an untyped memory area,
allocation is strictly delimited by the specified free memory
kernel objects are not shared or reused

This memory management model leaves the responsibility for verifying security policies outside the kernel. All that is left is to verify the correctness of the memory allocation algorithm in the kernel. The properties of interest being
that allocated objects are within the corresponding areas of free memory
and
that memory regions allocated to objects do not overlap.

Memory allocation capabilities can be transferred between the kernel components.
Transfers are represented as a tree in which the capabilities are the nodes of the tree.
Freeing memory is done in two steps that invalidate all references to that region:
search for all the capabilities for which access rights are granted on the memory object
and then
delete all these capabilities and mark the memory region as free.
For the first stage, the capability transfer tree is used to find and invalidate all capabilities that allow permissions on the memory region. In the second stage, it is verified through the same tree that there are no references in other objects or global references to the area to be released.

\subsection{Memory access verification}
Memory access is an interesting topic in order to model as accurately as possible the behavior of a C program.
In Isabelle pointers are represented as a new type of data, \code{datatype a ptr = Ptr word32}, which means that the pointer is represented only by the 32-bit address it contains. Using this representation one can reason about heap memory.\\
Here an important problem is raised when we pass from one pointer type to another.
For example,
if we have two \code{float} and \code{int} pointers to the same address,
after we use one to change the value from the address to which it points,
we cannot be sure that the other has not been changed. To ensure that pointers of different types point to different addresses,
the Burstall-Bornat model is used as a solution~\cite{burstall}
where heap memory is separated into types. 
Thus each data type has its own function that maps pointers to their values:
\begin{lstlisting}[escapeinside={(*}{*)}]
record state =
  heap_int :: word32 (*$\rightarrow$*) int
  heap_float :: word32 (*$\rightarrow$*) float
  heap_intptr :: word32 (*$\rightarrow$*) addr...
\end{lstlisting}

% This model avoids ambiguities when a pointer of a certain type is modified. In the previous example, if we changed the values contained in the address of an integer pointer, there is no confusion that a pointer to a float value was accidentally changed because the values of the float type are separated from each other. Of course, the fact that this model gives us the certainty that we do not accidentally change values of different types, however, also brings the limitation that it is no longer possible to switch from one data type to another by cast. A memory area, once allocated, remains defined in the corresponding heap memory section until it is released.

While this solves the issues mentioned above,
it also renders type casts unusable. A memory area, once allocated, remains defined in the corresponding heap memory section until it is released.

%--------------------------------------------------------------------------------------
\section{Methodology: C to Isabelle conversion}

A key component of the formal check in seL4 is the bridge between C language and the proofs in Isabelle~\cite{isabelle}.
This is also the most complex part of the proofs because the semantics of the C language must be taken into account such as the ones mentioned in the previous section but also data structures storage, pointer arithmetic and others.
In Isabelle memory addressing is represented by a function defined on the address space without information about the type of data to which the address refers.
The way different types of data are stored is treated separately~\cite{tuch}. Abstracting how memory access, data alignment, and how different data types are modeled removed the need for higher-level proofs to employ repetitive checks,
such as that a pointer is not null before being accessed,
instead these checks are already defined as constraints.

The correctness of the C language semantics is not, however, treated as critical to the proofs of the whole system because it adds an additional verification level:
the validation of the correspondence between the formal model and the result obtained after compilation.
% However,
% this validation brings the expected results when the code is compiled with a low level of optimizations. Therefore, the exclusive use of checking the resulting binary file without taking into account the source code was not completely performed.
The proof technique used to ensure the correspondence between the abstract specifications, the formal model of the source code,
and the model resulting from the analysis of the binary file 
is called a refinement.
A refinement is defined in~\cite{verifos} as: "Program C is a refinement of program A, if the set of behaviors of program C is a subset of the behaviors described by program A". 
Here a behavior means a sequence of steps given by a change of system state and the transition between these states. The state of the system consists of the state of its components (memory, processes, resources) belonging to the user space and to the kernel space.

%------------------------------------------
\subsection{Isabelle/HOL theorem prover}

Isabelle is an interactive theorem prover that supports several types of formal logic systems.
Isabelle/HOL is Isabelle's specialization of Higher Order Logic (HOL).
HOL is a type-based logic whose system resembles the one from functional programming languages~\cite{funcprog}.
Existing types can be classified~\cite{sel4} into:
\begin{itemize}
    \item basic types, e.g. bool(boolean), nat($\mathbb{N}$) or int($\mathbb{Z}$)
    \item type constructors, e.g. list and set types. Type constructors are written postfix, that is, after their arguments. For example, nat list is the type of lists whose elements are natural numbers.
    \item types of functions are denoted by "$\Rightarrow$";
    \item types of variables are denoted by $'a, 'b$, etc.
\end{itemize}
Terms are represented like in functional programming: by applying functions to certain types of arguments. If we have $f$ a function of type $\tau_1 \Rightarrow \tau_2$ and $t$ is a term of type $\tau_1$ then $f t$ is a term of type $\tau_2$. In Isabelle the notation $t::\tau$ is used to represent that the term $t$ is of type $\tau$.
Isabelle's proofs are structured in theories. A theory is a collection of types, functions and theorems, just like a module in a programming language. A theory has the following format
\begin{lstlisting}
theory T
imports B1 ... Bn
begin
  statements, definitions, proofs
end
\end{lstlisting}
where \code{B1 ... Bn} are the names of the existing theories on which the T theory is based. Each T theory must be in a file called \code{T.thy}.
HOL contains a Main theory, which contains all predefined basic theories, such as arithmetic, lists, or sets. A theory can include a list of more \code{.thy} files.
In practice, to have all theories needed for parsing and basic proofs we have to include \code{AutoCorres.AutoCorres}.
Proofs can take the form of theorems or lemmas, both can be used inside other proofs. There are specific keywords for applying these in order to reach our goal~\cite{isabelle}, for example the most common keywords used in our proofs are: \code{unfolding x} - which applies the definition of \code{x} on the current goal and \code{apply x} - which refers to other theorems or set of rules to be used.

%-------------------------------------
\subsection{Parsing C to Isabelle}

Approaching the C language from the perspective of obtaining a semantic model on which valid reasoning can be built is an important contribution of the seL4 system and deserves to be studied in detail. Several steps are taken to translate the C code from seL4 into Isabelle~\cite{greenaway}, steps that we will also need to take for the OpenBSD driver:
\begin{enumerate}
    \item each C source file is parsed by an external preprocessor, which extends \code{\#include} formulas and macro commands and other directives
    \item the result is translated into Simpl by the C-to-Isabelle analyzer~\cite{tuch}
    \item each structure in the program is represented by a record in Isabelle
    \item local and global variables are analyzed to generate two new types: a global variables record \code{globals}
    and \code{"a myvars"} record for locals
    \item functions are translated in equivalent Simpl language representation;
    \item proofs are performed on the generated functions to specify which global variables modify them
\end{enumerate}
The post-translation steps in Simpl are embedded in the AutoCorres tool~\cite{greenaway}.
Because this tool uses the result of the C-to-Isabelle parser as input,
AutoCorres supports the same subset of the C language.
Programs that use loops, function calls, cast between various types, pointer arithmetic, structures, and recursion are supported,
but references to local variables, "goto" and "switch" expressions, unions, floating point arithmetic operations or the use of pointers to functions
are not supported.
The example in \cite{greenaway} shows how one can go from the implementation in C of a simple function to the C-to-Isabelle parser output (with which it is quite difficult to work) and then to the final form after running the AutoCorres tool.
In essence, the purpose of the AutoCorres tool is to abstract the low-level representation from the C-to-Isabelle parser into a high-level one by:
\begin{itemize} 
    \item performing the conversion between the deeply embedded representation to the shallowly embedded one (as described below)
    \item abstracting the arithmetic operations at 32-bit machine word level into operations on the whole set of integers and natural numbers
    \item abstracting the heap memory at byte level into separate data-type areas using the Burstall-Bornat model~\cite{burstall}
    \item simplifying the code and translating the variable types from the Simpl representation into a form that is easy to reason in Isabelle.
\end{itemize}

\textbf{Deeply vs shalowly embedded representations.} Before we can begin to formally reason about a program, we must first translate it into the logic used by our theorem demonstrator. To bring the C code into Isabelle, it is necessary to decide which aspects of the code will be translated into the demonstrator logic.
If the emphasis is on the C program structure and its preservation in Isabelle,
we say that deeply embedded representation is used.
If the semantics of the program are important in the translation process,
then we have a shallowly embedded representation of the source code in Isabelle logic.
AutoCorres has the role of conversion between the structural representation of the C language given by the C-to-Isabelle parser into the semantic representation on which reasoning will be performed.

An example from \cite{greenaway} tries to explain the difference between the two forms of code representation starting from: $2 + 2 = 4$.
If we want to prove that the left side is equal to the right side,
we perform the addition (treating the expression as shallowly embedded) and state that the proposition is true.
If we look at the structure of the equation (deeply embedded),
on the left we have 3 characters and on the right only one.
Thus we can say that the two parts are not equal because we did not give any semantics to the assembly operation and its terms.
Structural treatment is not helpful if we want to prove certain statements about a program.
For this reason, the semantic representation of the C code is an important contribution in the verification of the seL4 kernel, and this is done through the AutoCorres tool.

The semantic representation obtained with AutoCorres aims to capture the behavior of C programs
where
the representation in Isabelle can show that the program
might change the overall state of the system,
might contain loops which may not end,
might have exceptions or other errors
and so on.
These requirements are covered by the extensive use of existing monads in Simpl (Skip, Basic, Cond, Guard, etc.) and the addition of new constructs such as \code{gets}, \code{return}, \code{whileLoop}. The later provides a great similarity between the imperative language of the source 
and the functional one in which it is modeled.

%--------------------------------------
\subsection{C subset limitations}
In our work we needed to tackle the C-language constraints mentioned above,
so we used only a subset of the C99 standard specifications~\cite{iso}.
The most relevant restriction is that pointers to functions are not supported.
Pointer data types are defined as functions that return data stored at those addresses. If the pointers refer to the address of a function, there is no guarantee that the reference cannot be circular and that the address of the function must also be resolved.
Other issues that we ran into include control flow sequences such as code jumps using "goto" or "switch" which are not supported and
compiler optimization for data positioning in memory when dealing with unions or bit fields.

%--------------------------------------
\textbf{Calling function pointers.}
The limitation of not being able to call functions that were set via their address to a function pointer was a major drawback in the integration of the OpenBSD driver because we needed to address programmable tasks to be executed in the future. The tasks may come from the device driver, the timer or other sources.
The main loop can only call the corresponding function that was set via its pointer.
Below we depict a simplified program to showcase the issue where the C-to-Isabelle parser fails to translate the last function because the call to \code{foo()} is done via the function pointer \code{p\_fun}. 
%-------------------------------------------------
\begin{lstlisting}
static int counter;
void foo(void) { counter++; }
void (*p_fun)(void);
void set_function(void) { p_fun = foo; }
void call_function(void) {
    if(p_fun) p_fun();
}
\end{lstlisting}
%-------------------------------------------------
Workarounds cannot provide the full proof, they only skip certain parts of the program or proofs in order to provide a translation avoiding the part were the function pointer is used.
We list here a few options:
\begin{itemize}
    \item skip parsing \code{call\_function} by adding the \code{DONT\_TRANSLATE} annotation, we used this in the proof because the other translations were not affected, we only had to avoid proving some properties that involved the function pointer;
    \item add the following adnotations before parsing the C file, this will assert those theorems as axioms rather than try to prove them:
\begin{lstlisting}
declare [[quick_and_dirty = true]]
declare [[sorry_modifies_proofs = true]]
\end{lstlisting}
    \item add annotations before parsing the C file, this will not try to prove the theorem that involves function pointers
    \begin{lstlisting}
[[calculate_modifies_proofs = false]]
\end{lstlisting}
\end{itemize}

%----------------------------------------------------------------------------------
\section{Driver verification\protect\footnotemark}

\footnotetext{Theorems, code and data
available at \url{https://gitlab.com/system.verification}
}

Drivers are pieces of software that are part of a monolithic kernel (but can also run is userspace), whose purpose is to interact with hardware devices or buses and to provide a interface between the kernel and those components.
We choose to verify drivers as a further development of seL4 verification because drivers are independent enough from the kernel structure, thus the verification process does not need to take into consideration the particularities of the kernel where the driver came from.

The main objective of driver modeling in Isabelle is to generalize the verification of kernel drivers and make it OS-agnostic.
We started from an OpenBSD driver which suffered adaptations meant to decouple its dependency on the kernel mid-layer.
This simulation comes at a cost,
we have to assume that the rest of the system works correctly because the verification will cover only the driver functionality.
We applied this assumption to hardware related components like
bus communication and reading/writing form device registers.
We assume that the bus works correctly and the register behavior matches the datasheet specifications.
In general,
this separation between the software driver and the hardware components
is helpful for identifying the source of defective device behavior.

%------------------------------------------
\subsection{OpenBSD octrng driver}
The driver used for prototyping seL4 verification is a hardware random number generator for Octeon boards.%~\cite{octrng}.
We choose this driver because it has a small configuration sequence
and
it is pretty isolated from the OpenBSD kernel
(there are no major dependencies from other drivers or kernel components).
The driver structure is very simple, it has two important functions.
First the driver initialization routine, \code{octrng\_attach}
whose purpose is to configure the hardware in order to start generating random values.
To do this it maps the registers of the device in the main address space and sets bits 62 and 63 (\code{OCTRNG\_ENABLE\_OUTPUT, OCTRNG\_ENABLE\_ENTROPY}) of register 0x1180040000000 (\code{OCTRNG\_CONTROL\_ADDR}).
The device starts generating random values.
% and its output is read every 10 milliseconds.
Afterwards,
\code{octrng\_rnd}, the second function, 
is called periodically to retrieve the random value generated by the device
from register 0x1400000000000 (\code{OCTEON\_RNG\_BASE + OCTRNG\_ENTROPY\_REG}).
The random value is be added to the entropy pool on each call.

%-------------------
\subsection{Mid-layer decoupling}

Before parsing the C driver implementation into Isabelle, 
some OpenBSD kernel mid-layer particularities had to be decoupled and
implemented separately so that the driver can stand on its own.
%We had four components that needed to be mimicked:
We mimicked:

\textbf{Bus communication.} The original driver accesses the bus via \code{bus\_space\_x()} functions, where \code{x} can refer to register mapping, reading or writing on the bus. In our case, we replace the bus access with simple reading or writing to local memory. This way, bus behavior is copied for read/write commands except for the timing (a bus write may need more time than writing to a local variable). In our case timing is not relevant because all actions are done sequentially.

\textbf{Device registers.} Because the bus communication is simulated,
we implement and express register behavior using local memory with a static structure containing only the required fields from the registers.
For octrng driver, we only need the control register, so we had a static structure \code{rng\_regs} with only one member \code{control\_addr} which will be the absolute address of the control register.
% \begin{lstlisting}[language=C]
% static struct reg {
%   unsigned long control_addr;
% } rng_regs;
% \end{lstlisting}

Reading and writing the device register is done by mapping the physical registers in memory.
This involves communication with the device via the bus on which it is located.
For our model however, the device is just a representation of the actual one, so there are no physical registers and our bus transfers are simply read-write operations from the device register structure.

Our model resembles as much as possible the internal register behavior.
For the octrng driver,
only some registers are important and so we have to cover only these cases: enabling the output bit, the entropy bit and reading the control value.
We implement this with two helper functions \code{set\_register} and \code{get\_register}.
The first function modifies the required register with a given value while the second one reads the control register or returns the value of the current timer if both output and entropy flags are set. \\
\textbf{Global timer.} In our model the timer serves two purposes.
The first is inherited from the original driver: scheduling a call to the random function every 10 milliseconds.
The other has been added for verification purposes and is not present on the actual hardware device: mimicking the random value by returning the timer value instead of the random value from the device register.
Note that because we do not have access to an actual timer,
we will simply use a global variable that will be incremented by the \code{idle()} function each time the main loop schedules a task (see below).

\textbf{Task scheduling.} The initialization call to the \code{attach} function of the driver is done from a separate file whose purpose is to simulate a very simple scheduler. The scheduler is a loop guarded by a timeout where we check for tasks waiting to be scheduled during each iteration. This loop also calls the \code{idle} function to increase the global timer. Tasks are stored into a static structure array whose members are the \code{timeout}, the \code{start} (or arrival) time
and the \code{timeout\_fun} callback.
% \begin{lstlisting}[language=C]
% typedef struct {
%     int timeout;
%     unsigned long start;
%     void (*timeout_fun)(void);
% } Task;
% \end{lstlisting}
Scheduling a task to run function \code{foo()} after 3 time units in the future implies adding a new task in the task queue with \code{timeout} set to 3, \code{start} set to the current time value and \code{timeout\_fun} pointing to the \code{foo()} function. The task queue is a circular buffer, each task addition increments the index of the newest task added. Tasks are removed from the buffer after completion.

%------------------------------------------------------------
\subsection{Proving driver function correctness in Isabelle}

We translate our driver model into Isabelle/HOL by applying successively the C-to-Isabelle parser and then the AutoCorres tool.
A limitation of these tools is that we can only parse one .c file at a time and provide one corresponding .thy file.
In seL4, some of the .c files have produced isolated theory files and these theories are then included where needed.
However, there is a starting point to parse all the other files and this is the \code{kernel\_C} preprocessor output file.
We used the same approach by including the \code{octrng} driver and the timer implementation inside the .c file containing the main loop.
% This file can be reprocessed by running the command 
% \code{gcc -DINCLUDE\_C\_FILES -E -CC run\_tasks.c > run\_tasks.c\_pp} and producing the \code{run\_task.c\_pp} file which will be used as input for the Isabelle translation.\\
% The \code{INCLUDE\_C\_FILES} compilation flag was added in order to easily switch between including the header and .c files into the main loop. The .c files are needed to be included instead of the headers because the otherwise we will not get the translation of the function implemented in the additional .c files. \\
The theory file contains the import statements that include AutoCorres theories and all the helper theories. The C-to-Isabelle parser is applied by declaring the input preprocessed file. After this step we have all the C functions translated into Simpl theorems. In order to obtain the final representation of these theorems, the AutoCorres tool is applied on the target file. Inside the main context of this theory we can start defining new terms, functions or proving new lemmas about the translated C functions.
% \begin{lstlisting}
% theory Run_Tasks
%   imports 
%     "AutoCorres.AutoCorres" 
% begin
%   external_file "run_tasks.c_pp"
%   install_C_file "run_tasks.c_pp"
%   autocorres "run_tasks.c_pp"
%   context run_tasks begin
%   definitions, lemmas, theorems ...
% end
% \end{lstlisting}

After the translation into Isabelle, we can access the functions from C as theorems in Isabelle. For example, C function \code{foo} is represented as a theorem named \code{foo’\_def}. 
%The definition of a function can be viewed in JEdit by typing $thm foo^\prime\_def$.
All additional functions implemented in all the included files will be translated. We analyze only the two functions related to the \code{octrng} driver: \code{octrng\_attach} and \code{octrng\_rnd}.
Any constants need to be redefined if we want to use the same names through the new theorems or lemmas. The C constants have been translated directly into their values, but we can give a name to the same values as Isabelle definitions (for example the enable output flag will be defined in Isabelle as \code{definition} \code{``OCTRNG\_ENABLE\_OUTPUT $\equiv (1 << 1) :: $ word32``)}.

\textbf{The attach function.} This is where the device configuration takes place and also the task of periodically checking the value is programmed. The resulting Isabelle translation of the associated modeled driver C code is:
\begin{lstlisting}[escapeinside={(*}{*)}]
Original:
    void octrng_attach(void) {
      unsigned long control_reg;
    
      control_reg = get_register(OCTRNG_CONTROL_ADDR);
      control_reg |= OCTRNG_ENABLE_OUTPUT;
      control_reg |= OCTRNG_ENABLE_ENTROPY;
      set_register(OCTRNG_CONTROL_ADDR,control_reg);
    
      add_task(octrng_rnd, 5);
    }
Isabelle:
    do ret' (*$\leftarrow$*) get_register' 0x0001180040000000;
      set_register' 0x0001180040000000 (ret' || 3);
      add_task' (PTR(unit) (symbol_table ''octrng_rnd'')) 5
    od
\end{lstlisting}
% \begin{figure}[htbp]
% \centerline{\includegraphics[width=0.4\textwidth]{attach_c.png}}
% \caption{\code{octrng\_attach} C implementation} 
% \label{fig:attach_c}
% \end{figure}
% \begin{figure}[htbp]
% \centerline{\includegraphics[width=0.45\textwidth]{attach_thy.png}}
% \caption{\code{octrng\_attach} Isabelle definition} 
% \label{fig:attach_thy}
% \end{figure}
We now want to verify that after the execution of \code{octrng\_attach} the device state is ready for generating random values, i.e. the control register is set corectly.
We model this inside a lemma in the form of a Hoare triple $\{P\} C \{Q\}$, where $P$ and $Q$ are the precondition and respectively the postcondition, C is the executed program.
In our case, we want to verify that running the $octrng\_attach$ program function in any program state, will result in the control register having set to 1  the enable output and entropy flags. So the precondition is always $True$ because there are no requirements and in the postcondition we check the bits of the flags.
\begin{lstlisting}[escapeinside={(*}{*)}]
lemma octrng_attach : "{| (*$\lambda$*)s. True |}
  octrng_attach'
{| (*$\lambda$*)_s.
   control_addr_C (rng_regs_'' s) && OCTRNG_ENABLE_OUTPUT (*$\neq$*) 0 (*$\wedge$*)
   control_addr_C (rng_regs_'' s) && OCTRNG_ENABLE_ENTROPY (*$\neq$*) 0 |} "
\end{lstlisting}
% \begin{figure}[htbp]
% \centerline{\includegraphics[width=0.45\textwidth]{attach_lemma.png}}
% \caption{\code{octrng\_attach} verification lemma} 
% \label{fig:attach_lemma}
% \end{figure}
This proof is straightforward, we only need to use \code{unfolding} to apply all the functions and definitions needed. The weakest precondition tool (\code{wp} command) computes the necessary precondition that we have to prove further. All the provided goals can be derived automatically from the function definition. Except for the bit operations where we need to explicitly apply the \code{word\_bitwise} theorems.

\textbf{The periodic “rng” function.}
This function should constantly retrieve the “random” value and add it to the pool. Because we only have the driver part and not the rest of the OpenBSD kernel, this value will be the timer value and the randomness pool will be just a global variable which will be updated by calling this function.
The modeled C implementation just reads the value from the output register and saves it into the \code{rand\_value} global variable,
then it schedules another function execution after 10 time units.
The Isabelle representation matches the same behavior,
the only difference is that all the global variables from the C program are now represented as Isabelle terms, for example the integer \code{rand\_value} is translated in Isabelle as \code{rand\_value\_''} a term of type sword32 (signed word on 32 bits).
\begin{lstlisting}[escapeinside={(*}{*)}]
Original:
void octrng_rnd(void) {
  unsigned int value;
  rand_value = get_register(OCTRNG_ENTROPY_REG);
  add_task(octrng_rnd, 10);
}
Isabelle:
do ret' (*$\leftarrow$*) get_register' 0;
  modify (rand_value_''_update ((*$\lambda$*)a. ret'));
  add_task' (PTR(unit) (symbol_table ''octrng_rnd'')) 10
od
\end{lstlisting}
% \begin{figure}[htbp]
% \centerline{\includegraphics[width=0.38\textwidth]{rnd_c.png}}
% \caption{\code{octrng\_rng} C implementation} 
% \label{fig:rnd_c}
% \end{figure}
% \begin{figure}[htbp]
% \centerline{\includegraphics[width=0.45\textwidth]{rnd_thy.png}}
% \caption{\code{octrng\_rng} Isabelle definition} 
% \label{fig:rnd_thy}
% \end{figure}
The verification lemma for \code{octrng\_rnd} has a few more preconditions than the initialization function because we have to first make sure that the function can be executed (the task queue is not full) and then that the driver is configured properly (the output and entropy flags are set). 
\begin{lstlisting}[escapeinside={(*}{*)}]
lemma octrng_rnd:
  "{| (*$\lambda$*)s. timer_" s = a (*$\wedge$*) running_tasks_" s < MAX_QUEUE (*$\wedge$*)
     current_tasks_" s < MAX_QUEUE (*$\wedge$*)
     control_addr_C (rng_regs_" s) && OCTRNG_ENABLE_OUTPUT (*$\neq$*) 0 (*$\wedge$*)
     control_addr_C (rng_regs_" s) && OCTRNG_ENABLE_ENTROPY (*$\neq$*) 0 |}
 octrng_rnd'
     {| (*$\lambda$*)_s. rand_value_" s=a |}! "
\end{lstlisting}
% \begin{figure}[htbp]
% \centerline{\includegraphics[width=0.45\textwidth]{rnd_lemma.png}}
% \caption{\code{octrng\_rng} verification lemma} 
% \label{fig:rnd_lemma}
% \end{figure}
The additional clause \code{$\lambda$ s. timer\_" s = a} represents that in any given state $s$ the global timer variable may have a label $a$ for its value. What we want to prove is that the same value will be set to the global \code{rand\_value} and this is the precondition \code{$\lambda$\_s. rand\_value\_" s = a}. The verification will be done using the same proofs as for the previous lemma: first we apply the definition of all functions used and then apply the weakest precondition tool. The goals obtained this way are easy to prove by applying the auto method.

This lemma could be improved by adding other specifications like checking that the same function will be called after 10 time units or that the function will be always called in time. The proofs that involve task scheduling were avoided because the translation of the function that runs the actual task is not parsed due to the issues described in the C subset limitation section.

\textbf{Main loop and other lemmas.}
The driver functions are bound together inside a small program that simulates a simple scheduler. The main loop does the initialization of the environment including the call of the \code{octrng\_attach} function,
then the main loop checks for each time unit if there are tasks whose timeout expired so their function has to be run.
We can add lemmas for those additional functions mainly because some of them might be useful in proving other properties.
For example, a simple function \code{idle} increases the global timer after each iteration of the main loop. The lemma for this function can verify that the timer is modified exactly by 1 after its execution in any program state.
\begin{lstlisting}[escapeinside={(*}{*)}]
lemma idle_increases [simp]:
  "{| s. timer_" s = a |}
  idle'
  {| (*$\lambda$*)_s. timer_" s = a + 1 |}! "
lemma main_function:
  "{| (*$\lambda$*)s. timer_" s = 0 (*$\wedge$*) running_tasks_'' s = 0 |}
  main'
  {| (*$\lambda$*)_s. timer_" s = TIMEOUT |}!"
\end{lstlisting}
% \begin{figure}[htbp]
% \centerline{\includegraphics[width=0.25\textwidth]{idle_lemma.png}}
% \caption{A lemma for \code{idle} function} 
% \label{fig:idle_lemma}
% \end{figure}
Its proof is obvious, we only have to apply the weakest precondition tool and then the \code{auto} method for applying the simplifications.
A proof that is more interesting is the one that states the main loop runs until a timeout occurs. This is done by limiting the timer with a maximum value, if this value is reached no other task will be called. 
% \begin{figure}[htbp]
% \centerline{\includegraphics[width=0.4\textwidth]{main_lemma.png}}
% \caption{A lemma for \code{main} function} 
% \label{fig:main_lemma}
% \end{figure}
The difference between this lemma's proof and the other is that here we have loops so we have to first provide a proof that those loops ends. Because the function that actually runs the task is not parsed, we will only prove the main loop, the one that increases the timer via the \code{idle} function and continuously run until timeout. This aspect is specified in the \code{main\_function} lemma: if we call the main function from a state where the timer is not started and there are no running tasks, then at the end the timer will have reached the timeout value. In order to prove this loop we have to specify and invariant and a measure.

The invariant is a property that has to be true before, during and after the main loop ends - because we want to prove something about the timer value, the invariant specifies that at any state of the loop, the timer will have a value between 0 and the timeout limit.
The measure is a value that has to decrease at each iteration - following the same model, the measure in our case is the distance between the timer and the timeout limit.
\begin{lstlisting}[escapeinside={(*}{*)}]
definition
 timer_limits_inv :: "word32 (*$\Rightarrow$*) 's lifted_globals_scheme (*$\Rightarrow$*) bool"
where
 "timer_limits_inv a s (*$\equiv$*) a = timer_" s (*$\wedge$*) 0 (*$\leq$*)timer_"  s (*$\wedge$*) 
  timer_" s (*$\leq$*) TIMEOUT "
definition
 timer_limits_measure :: " 'a (*$\Rightarrow$*) 's lifted_globals_scheme (*$\Rightarrow$*) word32"
where
 "timer_limits_measure a s (*$\equiv$*) a = TIMEOUT - timer_" s "
\end{lstlisting}
% \begin{figure}[htbp]
% \centerline{\includegraphics[width=0.45\textwidth]{invariant.png}}
% \caption{Definition of invariant and measure for main loop} 
% \label{fig:main_loop_lemma}
% \end{figure}
We can apply these two definitions via the \code{whileLoop\_add\_inv} monad and obtain a proof goal that can be further broken into smaller goals using the weakest precondition tool.

%---------------------------------------------------------------
\section{Conclusions}

In this paper we adapted and made use of the seL4 verification framework
to show that we can use the theorems and proofs of a micro-kernel operating system to successfully verify the octrng driver of the monolithic OpenBSD kernel.
Besides that,
we also provided a proof of concept regarding the verification of other mid-layer kernel components such as the scheduler.
While this is just a small part of the large OpenBSD code base,
our efforts lead to an encouraging conclusion: that the automatic abstraction of the source code using the AutoCorres tool reduces the complexity of the effort to demonstrate \cite{sweat} the properties of any system outside seL4.

We hope that in the future
this direction could facilitate the inclusion of verification as an important step in the development of system critical software. 

% These constraints seem to be the major barrier for expanding the C-to-Isabelle and AutoCorres tools into a generalized framework for safely translating C code into theorems.
% If the step of this translation could allow all programming language features to be used, some other C critical programs could be brought into Isabelle and reason about the correctness of their specifications.

%--------------------------------------------------------------------------------------
\clearpage

\bibliographystyle{splncs04}
\bibliography{bib}

%--------------------------------------------------------------------------------------
\end{document}